\documentclass[12pt]{iopart}
\usepackage{graphicx}
\usepackage{amssymb}
\usepackage{psfrag}

\makeatletter

\@addtoreset{equation}{section} \makeatother

\begin{document}
\title{\bf Coupling Hybrid Inflation to Moduli}

\author{Philippe Brax$^1$, Carsten van de Bruck$^2$,
Anne-Christine Davis$^3$ and Stephen C Davis$^4$}

\address{$^1$ Service de Physique Th\'eorique, CEA/DSM/SPhT,
Unit\'e de recherche associ\'ee au CNRS,
CEA-Saclay F--91191 Gif/Yvette cedex, France.}
\address{$^2$Department of Applied Mathematics, University of Sheffield
Hounsfield Road, Sheffield S3 7RH, United Kingdom}
\address{$^3$ DAMTP,Centre for Mathematical Sciences, University of
Cambridge, Wilberforce Road, Cambridge, CB3 0WA, UK}
\address{$^4$ Instituut-Lorentz for Theoretical Physics
Postbus 9506, NL--2300 RA Leiden, The Netherlands}

\eads{\mailto{brax@spht.saclay.cea.fr}, \mailto{C.vandeBruck@sheffield.ac.uk},
\mailto{A.C.Davis@damtp.cam.ac.uk} and \mailto{sdavis@lorentz.leidenuniv.nl}}

\begin{abstract}
Hybrid inflation can be realised in low-energy effective string
theory, as described using supergravity. We find that the
coupling of moduli to $F$-term hybrid inflation in supergravity
leads to a slope and a curvature for the inflaton potential. The
$\epsilon$ and $\eta$ parameters receive contributions at tree
and one loop level which are not compatible with
slow roll inflation. Furthermore the coupling to the moduli sector can
even prevent inflation from ending at all. We show that introducing shift
symmetries in the inflationary sector and taking the moduli sector to
be no-scale removes most of these problems. If the moduli fields are
fixed during inflation, as is usually assumed, it appears that viable
slow-roll inflation can then be obtained with just one fine-tuning of
the moduli sector parameters. However, we show this is not a reasonable
assumption, and that the small variation of the moduli fields during
inflation gives a significant contribution to the effective inflaton
potential. This typically implies that $\eta \approx -6$, although it
may be possible to obtain smaller values with heavy fine-tuning.

\end{abstract}

\maketitle

\section{Introduction}
The theory of inflation provides an explanation for many observed
properties of our visible universe, in particular its homogeneity
and isotropy (see e.g.~\cite{infla1} and~\cite{infla2} for reviews).
It also gives an explanation for the origin of small
density perturbations which eventually gave rise to the structures
in our universe. These aspects are well understood. However, from
the particle physics point of view, a suitable candidate for the
inflaton field has yet to be found~\cite{Lyth}. Among the many candidates,
hybrid inflation~\cite{hybridlind} is a particularly well motivated one,
especially in supersymmetric models~\cite{hybridini}--\cite{hybridend}.

There has been a lot of recent work investigating the possibility of realising
inflation within string theory (see e.g.~\cite{quevedo}--\cite{linderev}
for reviews). The most natural framework for brane inflation is the
brane-antibrane system where an attractive force leads to a
potential for the interbrane
distance~\cite{braneantibrane0}--\cite{braneantibrane3}.
Inflation ends with an open
string tachyonic instability similar to hybrid inflation. This
scenario has been extensively studied and leads to interesting
phenomena such as $D$-string formation at the end of inflation. One
particular issue in these models comes from the need for moduli
stabilisation during inflation. If this is not achieved, the
potential for the moduli is of runaway type destroying the existence
of slow roll inflation. This problem can be circumvented using the
KKLT scenario whereby the Kahler moduli are stabilised
once the complex moduli have been stabilised and non-perturbative
gaugino condensation has occurred on D7 branes~\cite{KKLT}. The coupling to
inflation has been studied in KKLMMT where a fine-tuning of the
inflaton superpotential has been advocated~\cite{KKLMMT}. One can also use  a
shift symmetry to alleviate the $\eta$ problem~\cite{shift1}--\cite{shift5}.
Another way of realising hybrid inflation in string models can be
obtained using the D3/D7 system, see
e.g.~\cite{kalloshbrane1}--\cite{cosmo}. In this case the end of
inflation happens at a lower scale than the Planck scale, when the charged
open string fields between the branes condense. The D3/D7 system can be
modelled using an $N=1$ supersymmetric $D$-term inflation
model~\cite{kalloshbrane2}.  The interbrane distance plays the role of
the inflaton and the charged open strings between the branes are the
waterfall fields. When the
supergravity corrections are neglected, i.e.\ in global
supersymmetry, and assuming that the Kahler moduli have been
stabilised, the inflaton direction is flat. Slow roll inflation is
driven by the one loop logarithmic corrections as the one loop
quadratic divergences are inflaton-independent.

In this paper, we are investigating $F$-term hybrid inflation with
the inflationary superpotential
\begin{equation}\label{inflasec}
W_{\rm inf}(\phi^\pm,\phi) = \sqrt 2 g\left( \phi^+ \phi^- - x^2
\right)\phi\, ,
\end{equation}
where $g$ is the $U(1)$ gauge coupling, in the presence of additional moduli
fields. It would be very interesting to derive the origin of
$x$ from fundamental string theory~\cite{koyama,keshav}.
We treat the supergravity case going beyond the global
supersymmetry analysis. We show that the treatment of Fayet-Iliopoulos
$D$-terms (i.e.\ when $x=0$) needs extra care in supergravity. To
include such a term would require additional fields and the study of their
stabilisation. For simplicity we therefore do not include inflationary
$D$-terms in our models. Inflation is then driven by the presence of the
non-vanishing $x$.

We show that supergravity corrections, and
supergravity-induced interactions with a moduli sector induce
a tree level slope and mass for the inflaton~\cite{Lyth}. Moreover there are
additional contributions from the quadratic divergence part of the
one-loop corrections, since they are inflaton-dependent in
supergravity, as opposed to inflaton-independent in global supersymmetry.
In general these effects spoil the flatness of the inflationary
potential. We show that most of the tree-level problems can be
removed by including shift symmetries in the inflaton sector, and
taking the moduli sector to be no-scale. It then appears that a viable
slow-roll inflation model can be found with limited fine-tuning of the
moduli sector. However we also find that the coupling between the
inflaton and the moduli has the effect of inducing a small variation
of the moduli during inflation. Despite the small size of this
variation, it gives a significant contribution to the inflaton slope
and violates slow-roll inflation.

The paper is arranged as follows. In section~\ref{sec:V}, we
construct the full potential for our combined model of inflation
and moduli stabilisation. We calculate tree and one loop level
parts of the potential, and describe symmetries which help
maintain the flatness of the inflaton potential. In
section~\ref{sec:mod} we determine the role of moduli fields
during inflation, and show their effect is significant. We find
the constraints corresponding to the COBE normalisation and the
WMAP3 spectral index constraint in section~\ref{sec:con}. Our
arguments and analysis apply to a general moduli sector, although
as a specific example, we also give explicit results for the KKLT
model, which has just one moduli field $T$. In
section~\ref{sec:kklt} we study particular choices of racetrack
superpotentials for the KKLT scenario, and find that the running
of the moduli field during inflation typically leads to a large
value of $\eta$. A small value of $\eta$ could only be achieved if
the moduli superpotential is heavily fine-tuned. Finally in
section \ref{sec:nogo} we consider models where a Minkowski vacuum
after inflation is obtained without the need for a lifting term. A
model with this property is also interesting as it may give a
small gravitino mass, as is required to have an low energy sparticle
spectrum. We derive a no-go theorem which states that, for
no-scale models with no lifting term, no such stable supersymmetry
breaking Minkowski vacua can be found. We conclude in section~\ref{sec:conc}.

\section{Combining hybrid inflation with moduli fields}
\label{sec:V}

As already discussed, many string theory inflation scenarios can be
modelled at low energy with a supergravity description of hybrid
inflation. The system uses three fields, the inflaton $\phi$ measuring
the inter-brane distance and two charged fields $\phi^\pm$ (which in
the D3/D7 system would represent the open strings between the two
types of branes). The fields interact according to the
superpotential~(\ref{inflasec}) with $x=0$, which has been studied
extensively in the literature. In~\cite{kalloshbrane2} it was shown
how the Fayet-Iliopoulos $D$-term arose from fluxes on the D7 brane,
allowing $D$-term inflation in this system.  Here we will focus almost
exclusively on hybrid inflation which is driven only by $F$-terms (so $x
\ne 0$), although exactly how such terms are embedded in string theory
remains to be found. Including inflationary $D$-terms in a supergravity
theory creates extra complications, as we will discuss in the
following subsection.

\subsection{Supergravity P-term inflation}

 Using~(\ref{inflasec}) and taking to $x \ne 0$ we obtain
\begin{eqnarray}
V &=& \left|\partial_i W_{\rm inf}\right|^2 + V_D \\ \nonumber
&=& 2 g^2 \left(|\phi\phi^+|^2 + |\phi\phi^-|^2 + |\phi^+ \phi^- -x^2|^2\right)
+\frac{g^2}{2}\left(|\phi^+|^2 - |\phi^-|^2 - \xi_1\right)^2 \, .
\label{susyinf}
\end{eqnarray}
This potential has an $N=2$ origin which was derived in~\cite{pterm}.
During inflation, when the waterfall fields $\phi^\pm=0$, the potential is
\begin{equation}
V = \frac{g^2\xi^2_1}{2}+2g^2 x^4 \, ,
\end{equation}
which gives a positive contribution to the inflation energy.

Extending this model to a supergravity theory, and working in units
with $M_\mathrm{Pl}=1$, we replace the above $F$-term by
\begin{equation}
V_F = e^{K_{\rm inf}}
(K^{i \bar \jmath} D_i W_{\rm inf} D_{\bar \jmath} \bar W_{\rm inf}
- 3|W_{\rm inf}|^2)
\end{equation}
where $D_i W_{\rm inf} = \partial_i W_{\rm inf}+ K_i W_{\rm inf}$, and
$K_{\rm inf}$ is the Kahler potential for $\phi$ and $\phi^\pm$. The
simplest possibility would be $K_{\rm inf}= |\phi|^2+ |\phi^-|^2+ |\phi^+|^2$.

It was originally thought that extending the $D$-term in the above
model to SUGRA would be straightforward. However it was later
discovered that in order to have a Fayet-Iliopoulos term $\xi_1$
appearing in a supergravity theory, the superpotential must have
charge $-\xi_1$~\cite{pierre}. Thus the extension of~(\ref{susyinf})
to SUGRA described in the earlier paper~\cite{pterm} requires
reconsideration. To preserve the properties of $\phi$ and its role
as the inflaton, it should be uncharged~\cite{pierre}.  This is only
possible when $x=0$ and the charges of $\phi^\pm$ are shifted to
$q_\pm= \pm 1 - \rho_\pm\xi_1$, where $\rho_++\rho_-=1$. Now all the
contributions to $W$ from  any other sectors of the theory would
also have to be charged. This would be for instance the case with the
MSSM superpotential.  Since they are, {\em a priori}, unrelated to
inflation, this seems unnatural. Hence we have an incompatibility
between a non-vanishing $x$ and a Fayet-Iliopoulos term $\xi_1$. In
the usual inflationary scenario, inflation is driven by the
Fayet-Iliopoulos term.

One possibility is to introduce a Fayet-Iliopoulos term with a
gauge invariant superpotential. This can be achieved in
supergravity with a $D$-term by making the
$U(1)_1$ pseudo-anomalous. We extend the model to include one
modulus $X$ and matter fields $A^a$, all of which are charged
under the pseudo-anomalous $U(1)_1$ gauge group. The relevant part
of the Kahler potential for these fields is
\begin{equation}
K(X,A^a)= -\log \left( X+\bar X - \xi_1 \mathcal{V}_1 \right) +
\sum_i e^{-q_a \mathcal{V}_1} |A^a|^2 \, ,
\end{equation}
where the gauge field, $\mathcal{V}_1$ has been included (the usual
Kahler potential appearing in the scalar potential expression is
obtained by putting $\mathcal{V}_1=0$ formally).

The potential arising from the $D$-term is then
\begin{equation}
V_D= \frac{g^2}{X+\bar X} \left( \vert \phi^+\vert^2 - \vert \phi^-\vert^2
-  \frac{\xi_1}{X+\bar X} + \sum_a q_a \vert A^a|^2  \right)^2 \, .
\end{equation}
where we have taken the gauge coupling function to be $f=X/g^2$
in order to cancel the $U(1)_1$ anomaly. We see that
a Fayet-Iliopoulos term can give rise to inflation when $x=0$,
assuming that $X$ and $A^a$ can be stabilised by a suitable superpotential.

During inflation, the waterfall fields $\phi^\pm$ vanish.
This implies that the $U(1)_1$ $D$-term and the $x$ dependent part
of the $F$-term potential give rise to the inflation energy
\begin{equation}
V_0= \frac{g^2}{(X+\bar X)^3}
\left(\xi_1 - (X+\bar X)\sum_a q_a |A^a|^2 \right)^2
+\frac{2g^2 x^4}{(X+\bar X)^3} e^{\sum_a |A^a|^2} \, .
\end{equation}
When $x=0$ and the extra fields are stable, this resembles the usual inflation
energy in hybrid inflation.

If on the other hand we just use the $F$-term to give inflation,
there is no need to introduce the additional fields $A^a$ and $X$.
The Fayet-Iliopoulos $D$-term must then be zero, due to the gauge
invariance of $W$, so $\xi_1=0$. We will take $x\ne 0$ and
assume this is the case for the rest of the paper. The string theory origin 
for $x$ has yet to be found~\cite{keshav}.

\subsection{Combined tree-level potential}
\label{ssec:VV}

We are interested in the hybrid inflation scenario based on the
superpotential above~(\ref{inflasec}). We are also taking into
account the effects of a separate sector, so the total
superpotential is
\begin{equation}
{\cal W}(Z^I,\phi,\phi^\pm)  = W_{\rm inf}(\phi,\phi^\pm) +
W(Z^I) \, .
\label{Wfull}
\end{equation}
The $Z^I$ are moduli and matter fields, which are usually assumed to
be stable during inflation. With one modulus in $F$-theory no
stabilisation was obtained~\cite{sethi}. Here we consider the most
general case in supergravity with more than one moduli field and
generic superpotential $W(Z^I)$. The moduli dependent part of the
superpotential $W(Z^I)$ is then responsible for stabilising the
moduli fields.

The total Kahler potential has the form
${\cal K} = K_{\rm inf} + K(Z^I)$ with
\begin{equation}
K_{\rm inf} = -\frac{1}{2}(\phi - \bar\phi)^2 +(\phi^+
-\bar\phi^-)(\bar\phi^+ - \phi^-).
\end{equation}
We have imposed two shift symmetries in the $(\phi,\phi^\pm)$
sector to alleviate the $\eta$ problem of supergravity
inflation. This is an extension to the usual $\phi$ shift
symmetry~\cite{shift1}--\cite{shift5}. In string theory, this
corresponds to the translational invariance of the brane system.
For the full superpotential~(\ref{Wfull}), the $F$-term part of the
potential can be expanded as
\begin{eqnarray}
V_F = e^\mathcal{K} &\Big[& \left| \sqrt{2}g\phi \phi^- +
\left(\bar\phi^+ - \phi^- \right)\left(W_{\rm inf} + W \right)
\right|^2
\nonumber \\
&&{}+ \left| \sqrt{2}g\phi\phi^+ + \left(\bar\phi^- -
\phi^+\right)\left(W_{\rm inf}+ W \right) \right|^2
\nonumber \\
&&{}+ \left| \sqrt{2}g(\phi^+ \phi^- - x^2)
+ (\bar\phi - \phi)(W_{\rm inf} + W)\right|^2 \nonumber \\
&&{}+ V_2 |W_{\rm inf}|^2 + 2{\Re}\{V_1 \bar
W_{\rm inf}\} + V_S  \Big] \, ,
\end{eqnarray}
where $\Re(z)$ denotes the real part of $z$, and we have defined
\begin{eqnarray}
&& V_S = ( K^{I\bar J}D_IW D_{\bar J}\bar W - 3|W|^2) \, ,
\\ \nonumber &&
V_1 = (K^{I\bar J} D_IW K_{\bar J} - 3W) \, , \qquad
V_2= (K^{I\bar J}K_I  K_{\bar J} - 3) \, .
\end{eqnarray}
There is also a contribution to $V$ from the $D$-term
\begin{equation}
V_D^{(1)} = \frac{1}{2 \Re [f(Z^I)]}\left(|\phi^+|^2 - |\phi^-|^2\right)^2
\, .
\end{equation}

In this paper we will be particularly interested in Kahler potentials
with the no-scale property, such as $K=-3\ln(T+\bar T)$ in the KKLT
scenario. In this case the above expressions reduce to
\begin{equation}
V_2(Z^I)=0 \, , \qquad
V_1(Z^I)= K^{I\bar J} \partial_I W K_{\bar J} \, .
\end{equation}

At the end of inflation, the charged fields $\phi^\pm$ condense
and the inflaton $\phi$ is zero. The $U(1)_1$ $D$-term part of the
potential is then zero, and $F$-term part reduces to $V_F =
V_S(Z^I)$. This part of $V_F$, which comes from the moduli
dependent part of the superpotential $W(Z^I)$, is then responsible
for stabilising the moduli fields. For $W\ne 0$ its minimum is an
${\rm AdS}_4$ vacuum where supersymmetry is preserved, i.e.\ $F_I=0$ for
all the fields. Since we want a Minkowski vacuum at the end of
inflation, it is necessary to add an additional lifting term,
$V_\mathrm{lift}(Z^I)$ to the potential.  In the KKLT scenario  a
non-supersymmetric lifting term is used
\begin{equation}
V_{NS}=\frac{E}{(T+\bar T)^n} \, ,
\label{NSlift}
\end{equation}
where $n=2,3$ depending on the origin of the term, i.e.\ $n=2$ for
anti D3-branes and $n=3$ for fluxes on $D7$ branes.

An alternative possibility is that the minimum is raised by an uplifting
$D$-term~\cite{burgess,ana}. This allows one to lift the ${\rm AdS}_4$
vacuum while avoiding the need for a non-supersymmetric potential. In
this case the set of fields $Z^I$ comprises one modulus $T$ and matter fields
$\chi^i$, all of which are charged under a second, pseudo-anomalous
$U(1)_2$ gauge group. The Kahler potential in the moduli sector can
be written as
\begin{equation}
K(Z^I)= -3\ln \left( T+\bar T -\frac{\xi_2}{3} \mathcal{V}_2 \right)
+ \sum_i e^{-\tilde q_i \mathcal{V}_2} |\chi_i|^2 \, .
\end{equation}
Unlike the KKLT case it is not no-scale, although it is
straightforward to write down a no-scale version.

The potential arising from the $D$-term for the additional $U(1)_2$ is then
\begin{equation}
V^{(2)}_D \propto \frac{1}{T+\bar T} \left(\sum_i \tilde q_i |\chi_i|^2
+ \frac{\xi_2}{T+ \bar T}\right)^2 \, .
\label{Dlift}
\end{equation}
If $\xi_2$ and all $\tilde q_i$ are positive, this potential will
be nonzero and provide a suitable lifting term. As explained
in~\cite{Nilles}, an ${\rm AdS}_4$ vacuum in supergravity cannot
normally be lifted by $D$-terms. One can circumvent this argument by
using non-analytic superpotentials with positive definite $D$-terms,
as described in~\cite{ana}.

The total moduli stabilisation potential is then
\begin{equation}
V_{\rm stab}(Z^I) = e^K V_S(Z^I) + V_\mathrm{lift}(Z^I) \, ,
\end{equation}
where $V_\mathrm{lift}$ is given by, for example, (\ref{NSlift})
or (\ref{Dlift}). After inflation, the  potential possesses a Minkowski vacuum
thanks to the presence of the lifting term.

For the rest of the paper we will only concentrate on the case
where there are no additional matter fields $\chi^i$. This implies
that the gauge invariance of the superpotential can only be
maintained if $T$ is gauge invariant, and we must use the
non-supersymmetric lifting term~(\ref{NSlift}). This is the
minimal setting to study hybrid inflation coupled to moduli.

\subsection{Inflation}

The structure of this potential is rather rich due to the
possible interplay between the moduli and other fields.
During hybrid inflation, the fields $\phi^\pm$ vanish, and the
inflaton field is real, so $\bar \phi = \phi$. The full (tree-level)
potential then reads
\begin{equation}
V = V^\mathrm{(tree)}_{\rm stab} + V_0^\mathrm{(tree)} -
2\sqrt{2}gx^2 e^K \Re\{V_1\} \phi + 2 g^2 x^4 e^K V_2 \phi^2 \, ,
\label{inflapotential}
\end{equation}
where we have identified
\begin{equation}
V^\mathrm{(tree)}_{\rm stab}(Z^I)
= e^K V_S(Z^I) + V_\mathrm{lift}(Z^I)
\end{equation}
and
\begin{equation}
V_0^\mathrm{(tree)}(Z^I)  = 2 g^2 x^4 e^K \, .
\end{equation}
In addition to the terms above, there will be contributions from
loop corrections, which we will discuss in the next subsection.
Note that the inflaton is much smaller than the Planck mass, i.e.\
$\phi = \bar\phi$ is such that $\phi\ll M_{\rm pl}$. During
inflation, the Minkowski vacuum is lifted by $V_0$. We see that in
general we will get large, tree-level contributions to the slope and
mass of the inflaton~\cite{Lyth} (from $V_1$ and $V_2$ respectively). In other
words the model has $\epsilon$ and $\eta$ problems. This new $\eta$ problem
comes from the inclusion of the moduli sector, and is distinct from the
usual $\eta$ problem which comes from embedding inflation in
supergravity. In the case of no-scale models $V_2=0$, and so this
$\eta$ problem can be removed by a symmetry of the Kahler potential
$K$. To obtain a small inflaton slope, we will need to fine tune $V_1$.

The form of the potential discussed so far was based on the hybrid
inflationary epoch, in which the fields $\phi^{\pm}$ vanish.
Inflation ends when the mass matrix acquires a tachyonic
direction. This usually occurs when one of the two charged fields
$\phi^\pm$ (or a combination thereof) has a negative mass, i.e.\ when
$m^2_-<0$. Without moduli fields, this instability occurs at
$\phi=x$. However, the inclusion of the moduli sector modifies this
result. For simplicity  we take $V_2=0$ in what follows (as is the
case for a no-scale $K$). Working out the mass matrix for the
$\phi^\pm$ sector, we find
\begin{equation}\label{phiend}
\phi_{\rm end}= \sqrt{x^2 + \frac{V_1^2}{8g^2}}
- \frac{V_1}{2\sqrt{2}g} \, ,
\end{equation}
where we have assumed $W$ to be real. If $V_1=0$, we recover the
standard result $\phi=x$. However, due to the presence of the
moduli fields, the value of $\phi_{\rm end}$ is altered. Note that
if we had used a more general $K_{\rm inf}$, which did not have
the shift symmetries, then $\phi_{\rm end}$ would have received
far more corrections. Taking a canonical Kahler potential, and
neglecting the contributions of order $V_1$ in the regime
$x \ll 1$, the lowest mass is given by
\begin{equation}
m^2_-= e^K\left[
2g^2 (\phi^2 -x^2) + W^2 + 2\sqrt{2} g W\phi \right]\, .
\end{equation}
We see that if $W \geq \sqrt{2}g x$, inflation never ends as the
mass is always positive. We naturally expect that  $W$  will be much
larger than the inflation scale $x$ in order to stabilise moduli
during inflation. We see that both shift symmetries must be included
in $K_\mathrm{inf}$ for inflation to be viable.

In the above discussion we have considered how the moduli sector
will alter inflation. Of course we must not forget that the
inflationary sector can also interfere with the moduli
stabilisation, since the inflation potential $V_0$ acts as a
perturbation of the moduli potential $V_{\rm stab}$. In the KKLT
scenario the minimum of $V_\mathrm{stab}$ is separated from
$T=\infty$ (which also has $V_\mathrm{stab}=0$) by a potential
barrier. Denoting the height of the barrier by $V_\mathrm{max}$, we
see that the value of the above potential~(\ref{inflapotential})
must be less than $V_\mathrm{max}$ during inflation, or the modulus
field $T$ will roll off to infinity, and inflation will not finish
in the correct vacuum. The slope terms in the above potential must
be small (in order to have slow-roll inflation), so we need $e^K g^2
x^4 \lesssim V_\mathrm{max}$. This guarantees that the minimum of
the moduli potential is only shifted during inflation and not
completely destroyed.

Typically $V_\mathrm{max} \sim -V_{\rm AdS} \sim 3 W^2 e^K$, where
$V_{\rm AdS}$ is the energy of the AdS minimum of $V$ (if
$V_\mathrm{lift}$ were absent). This provides a rough estimate of
the barrier height. In this case we need $gx^2 \lesssim W$, which is
reasonable if the inflation scale is far below the moduli stabilisation scale.

\subsection{Loop corrections}
The one-loop corrections to the effective potential for a general
theory, with cut-off $\Lambda$, are~\cite{loop}
\begin{equation}
V_\mathrm{loop} = \frac{1}{32 \pi^2} \mathrm{Str} M^2 \Lambda^2
+\frac{1}{64\pi^2} \mathrm{Str} M^4 \log \frac{M^2}{\Lambda^2} \, .
\end{equation}
The supertrace is $\mathrm{Str} M^2 = M^2_{\rm (boson)} - M^2_{\rm
(fermion)}$. Expressions for the boson and fermion mass matrices
are given in the appendix. As long as the masses are below the
cut-off scale, the log corrections are smaller than the quadratic
divergences. This is guaranteed provided the potential $W(Z^I)$
is lower than the cut-off, i.e.\ lower than the Planck scale in
practice.  This is also a phenomenological requirement in order to
obtain a hierarchy between the Planck mass and the gravitino mass.
We therefore concentrate on the quadratic divergences in what follows.

We see there are corrections to all parts of the inflationary
potential~(\ref{inflapotential}). In particular we find
\begin{equation}
\tilde V_0 = c_* V_0 = 2g^2x^4 \left(1
+ \frac{\Lambda^2}{16\pi^2} \left[2+\delta^I_I\right]\right) e^K \, .
\end{equation}
In general, the corrections to the other parts of the potential are
complicated, and not particularly illuminating. For simplicity we
will explicitly consider the case for one modulus field only, i.e.\
$W=W(T)$. We will use $f(T)\propto T$. In the Minkowski background
\begin{equation}
\mathrm{Str} M^2_\mathrm{Mink} = \frac{2n(n+1) E}{3 (2T)^n}
+ \frac{ 2W'^2}{3T} - \frac{2WW'}{T^2} -\frac{W^2}{2T^3} \, .
\label{StrMink}
\end{equation}
During inflation we have instead
\begin{equation}
\mathrm{Str} M^2_\mathrm{Inf} - \mathrm{Str} M^2_\mathrm{Mink}
= \frac{g^2 x^4}{2T^3} (3 -2 \phi^2)
+ \frac{2\sqrt{2} gx^2}{T^3} (T W'+W) \phi \, .
\label{StrInf}
\end{equation}
We see that the loop corrections are $\phi$ dependent. Note that we
have included not only the contributions from $\phi^\pm$, but also
those from the moduli sector.

For the KKLT scenario, we find the tree-level potential to be (after taking
$T$ to be real)
\begin{equation}
V_{\rm tree}^{(\rm Mink)} = \frac{W'(TW'-3W)}{6T^2} + \frac{E}{(2T)^n}\, ,
\end{equation}
where $W' = dW/dT$. Adding on the loop corrections~(\ref{StrMink})
gives the full potential for the Minkowski background
\begin{eqnarray}
\tilde V_\mathrm{stab} &=&
\frac{W'^2}{6T}\left[1+\frac{\Lambda^2}{8 \pi^2}\right ]
- \frac{W W'}{2T^2}\left[1+\frac{\Lambda^2}{8 \pi^2}\right]
-\frac{W^2}{T^3} \frac{\Lambda^2}{32 \pi^2}
\\ \nonumber && {}
+\frac{E}{(2T)^n}\left[1+\frac{\Lambda^2 n(n+1)}{48 \pi^2}\right] \, .
\label{kkltVS}
\end{eqnarray}
Adding the loop corrections will change the value of $T$ at the
minimum. Furthermore the minimum will no longer be Minkowski. The
parameter $E$ needs to be re-tuned to fix this.

During inflation, i.e.\ $\phi$ non-vanishing, we get
\begin{equation}
V_{\rm tree}^{\rm (Inf)} = V_{\rm tree}^{\rm (Mink)}
+ \frac{\sqrt{2}x^2gW'\phi}{2T^2} + \frac{x^4g^2}{4T^3} \, ,
\end{equation}
to which we need to add the loop corrections~(\ref{StrInf}).

In this paper, we take the cut off at the Planck scale. Hence, as
long as $W'$ and $W$ do not vanish during inflation, the radiative
corrections induce a slope for the inflationary potential, which is
dependent on $W$, and which can in principle be a danger to the
flatness of the inflationary potential. This effect is only present
in supergravity, as in global supersymmetry the quadratic
divergences are $\phi$ independent. For one modulus field $T$, the
full potential during inflation now reads
\begin{equation}
V_{\rm total}^{\rm (Inf)} = \tilde V_{\rm stab} + \tilde V_0
- 2\sqrt{2}gx^2 e^K\Re\{\tilde V_1\}\phi + 2 g^2 x^4 e^K \tilde V_2 \phi^2 \, .
\label{VInf}
\end{equation}
For $T$ real, the full moduli stabilisation potential is given
above~(\ref{kkltVS}). We denote the value of $T$ at the minimum of $\tilde
V_\mathrm{stab}$ by $T_0$. Similarly
\begin{equation}
\tilde V_1 = -2 T W'\left[1+\frac{\Lambda^2}{8
\pi^2}\right] - W \frac{\Lambda^2}{4 \pi^2}
\label{kkltV1}
\end{equation}
\begin{equation}
\tilde V_2 = -\frac{\Lambda^2}{4\pi^2} \, .
\label{kkltV2}
\end{equation}
As it is a no-scale model, the only contribution to $\tilde V_2$
comes from loop corrections.

\section{Moduli fields during $F$-term hybrid inflation}
\label{sec:mod}

In this section we will discuss the form of the effective inflationary
potential, taking into account the moduli stabilisation mechanism. It
is usually assumed that moduli fields are stabilised before inflation
begins. The energy scale of the corresponding potential is much higher
than that of inflation, and so the effects of the inflaton will not
significantly alter the values of the moduli fields.  It therefore
seems reasonable to assume the moduli are fixed during inflation. As
we will show, this is not the case.

Let us now analyse the moduli stabilisation mechanism when the
inflation sector is present. We denote the values of the moduli
fields at the minimum of $\tilde V_{\rm stab}$ by $Z^I_0$, and define
$\delta Z^I = Z^I - Z^I_0$ to be the deviation of the moduli fields
from this value. Expanding the inflationary potential around this
minimum we find
\begin{eqnarray}
V(\phi,Y^a) &=& \frac{1}{2}\sum_a \lambda_a (Y^a)^2 -2\sqrt{2}g x^2
\phi \sum_a \Re (e^K \tilde{V}_1)_{,a}Y^a
\nonumber \\ && {}
+ \tilde V_0-2\sqrt{2}g x^2 \phi e^K \Re (\tilde{V}_1) + 2 g^2 x^4
\phi^2 e^K\tilde{V}_2
\nonumber \\ && {}
+\mathcal{O}(Y^3, Y^2 g x^2 \phi, Y g^2 x^4) \, ,
\label{VY}
\end{eqnarray}
where $Y^a$ are independent combinations of $\delta Z^I$ and $\delta \bar Z^I$
which satisfy
$\sum_b \tilde V_{\mathrm{stab},ab}(Z^I_0) Y^b$ $ = \lambda_a(Z^I_0) Y^a$.
Minimising this with respect to $Y^a$, we find
\begin{equation}
Y^a = 2\sqrt{2}g x^2 \phi \frac{\Re (e^K \tilde{V}_1)_{,a}}{\lambda_a} +
\mathcal{O}(g^2 x^4) \, . \label{Yeq}
\end{equation}
So we see that the moduli fields are now stabilised along a
valley depending on $\phi$. If $\lambda_a \gg g x^2 \phi$
(which is perfectly reasonable), then $\delta Z^I \ll Z^I_0$,
and the variation of the moduli fields is tiny, as is usually assumed.
The correction to the inflationary potential is tiny too. However
the slope of $V$ is also tiny, and so the contribution of the moduli fields
can be comparable to the other effects we are considering. Hence it is
not reasonable to ignore their variation during inflation after all.
Now if we substitute the above expression~(\ref{Yeq}) back into the
potential~(\ref{VY}), we obtain the effective inflationary potential
\begin{equation}
V(\phi)= \tilde V_0  + C \phi + \frac{M^2}{2} \phi^2
+ \mathcal{O}(g^3 x^6) \, ,
\label{Veff}
\end{equation}
with slope
\begin{equation}
C= -2\sqrt{2}g x^2 e^K \Re (\tilde{V}_1) \, ,
\end{equation}
and effective mass
\begin{equation}
M^2 =4 g^2 x^4 \left[ e^K \tilde{V}_2 - 2\sum_a
\frac{(\partial_a\Re(e^K \tilde{V}_1))^2}{\lambda_a}\right] \, .
\end{equation}
We remind the reader that loop corrections are included in these
expressions. Hence, we see for a wide range of models (including the
KKLT scenario) that $M^2<0$, and so the inflaton is tachyonic.  This
tachyonic instability is not necessarily a problem if $\phi$ is small,
or the mass term is subdominant. Since we need the field to roll
towards zero during inflation we also require that $C$ is positive.

It is tempting to use a moduli sector whose superpotential satisfies
$W=\partial_I W =0$ at the minimum, since $\tilde V_1 =0$ in this
case, and so we automatically avoid any large contributions to the
slow-roll parameter $\epsilon$. This type of model has other appealing
features, such as a small gravitino mass~\cite{kalloshlinde}. If we
consider just one moduli field $T$, and take a no-scale $K$, then we see
from (\ref{kkltV2}) that loop corrections give a small
negative inflaton mass. The effects of the moduli variation will make
it even more negative. Even if this mass is small, it will still
dominate the behaviour of $\phi$, and cause it to roll away from
zero. Hence we see that a moduli potential with $W= W' =0$ at its
minimum (i.e.\ unbroken supersymmetry) cannot be successfully combined
with $F$-term hybrid inflation.

\section{Slow-roll inflation constraints}
\label{sec:con}

We have established that during inflation the effective potential
has the leading order form
\begin{equation}\label{workingmodel}
V(\phi) = \tilde V_0 + C\phi + \frac{M^2}{2}\phi^2 \, ,
\end{equation}
with $C$ positive and $M^2$ negative. The potential has the form
of an inverted parabola. Since we want inflation to end by the
tachyonic instability of the $\phi^{-}$-field, $\phi$ has to roll
towards zero. This means that during inflation $\phi < \phi_c
\equiv -C/M^2$. Thus, an additional constraint on the model
parameters is that $\phi_c > \phi_{\rm end}$, as given
in~(\ref{phiend}). To get an idea about the allowed parameter ranges
for the theory, making only modest assumptions about the moduli sector, we
provide some general constraints on potentials of the
form~(\ref{workingmodel}), leaving $\tilde V_0$, $C$ and $M^2$ as free
parameters, but with $M^2$ negative. Note that $\tilde V_0$, $C$
and $M^2$ depend on the underlying theory and are {\it not}
independent. As shown in the previous section, they depend on the
details of the moduli sector as well as the parameters $g$ and
$x$. An inflationary model has to satisfy certain constraints in
order to agree with observations. Firstly, in order to realise a
period of slow-roll inflation, the two slow-roll parameters
\begin{equation}
\epsilon = \frac{1}{2}\left(\frac{V'}{V}\right)^2
\approx \frac{1}{2}\left(\frac{C + M^2 \phi}{\tilde V_0}\right)^2
= \frac{C^2}{2\tilde V_0^2}\left(1-\phi/\phi_c\right)^2
\end{equation}
and
\begin{equation}
\eta = \frac{V''}{V} \approx \frac{M^2}{\tilde V_0}
\end{equation}
have to be small (we have approximated $V\approx \tilde V_0$ during inflation).

Also, in order to get the right amplitude for the density
perturbations from inflation, the COBE normalisation must hold~\cite{Lyth}
\begin{equation}\label{COBE}
\frac{V}{\epsilon}\equiv \delta
\approx 24 \pi^2 (5 \times 10^{-5})^2 \, .
\end{equation}
The current WMAP+2dFGRS constraint on the spectral index
is~\cite{WMAP3}
\begin{equation}\label{nsconstr}
n_s = 0.948^{+0.014}_{-0.018}~.
\end{equation}
(a weaker bound has been found in~\cite{kolbinf,jerome}). In the
case that $\phi\ll\phi_c$, the above two constraints are fulfilled
when
\begin{equation}\label{norm}
\frac{2\tilde V_0^3}{C^2}\approx 24\pi^2 (5\times 10^{-5})^2 \, ,
\end{equation}
and
\begin{equation}
n_s - 1 = 2\eta - 6\epsilon = \frac{2M^2}{\tilde V_0} -
\frac{3C^2}{\tilde V_0^2} \, .
\end{equation}
The constant $C$ can be eliminated from the last expression,
using~(\ref{norm}). Since $M^2$ is negative and $n_s \approx 0.95$,
this restricts the parameters to be in the region (in Planck units)
\begin{eqnarray}
&0& \leqslant \tilde V_0 \lesssim 7\times10^{-9}\, , \label{V0con} \\
&0& \geqslant M^2 \gtrsim -6\times 10^{-11}\, ,\\
&0& < C \lesssim 10^{-9} \, ,
\end{eqnarray}
implying an upper bound for the inflationary scale of order
$2\times 10^{16}$ GeV. The allowed region in the $\tilde V_0-M^2$
plane is shown in Figure~1.
\begin{figure}
\psfrag{xl}[][][2]{$\tilde V_0$}
\psfrag{yl}[][][2]{$M^2$}
\centering
\scalebox{0.45}
{
 \includegraphics{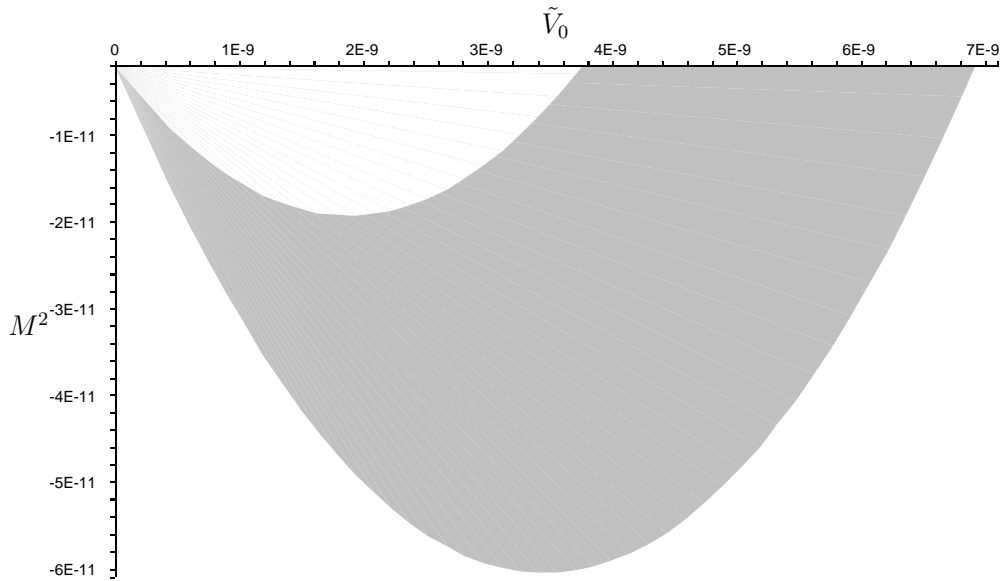}
} \caption{Constraints on the parameters $M^2$ and $\tilde V_0$
coming from the observed amplitude of the spectrum~(\ref{norm}) and
the observed spectral index $n_s$~(\ref{nsconstr}). The allowed
region is the grey area and the values are given in Planck units.
It is assumed that $M^2$ is negative, as predicted in these theories.}
\end{figure}

The parameters $V_0$, $C$ and $M^2$ depend on the details of the
moduli superpotential as well as on $g$ and $x$. In order to
generate slow-roll inflation, we require that
\begin{equation}\label{cb1}
\epsilon \approx \frac{C^2}{2\tilde V_0^2}
= \left(\frac{\Re (\tilde{V}_1)}{g x^2 c_*}\right)^2 \ll 1
\end{equation}
and
\begin{equation}
|\eta| \approx \left| \frac{M^2}{V_0c_*} \right|
=\frac{2}{c_*}\left|\tilde{V}_2 - 2\sum_a
\frac{(\partial_a\Re(e^K \tilde{V}_1))^2}{e^K \lambda_a} \right|
\ll 1 \, .
\end{equation}
Note that if we had not included the contribution from the variation
of the moduli fields, we would instead have $|\eta| =
(2/c_*)|\tilde{V}_2| \ll 1$. In general, the above contributions to
$\eta$ from the moduli sector are likely to be large. We see this
type of model may have another $\eta$ problem, which is not removed
by the symmetries we used in section~\ref{sec:V} to avoid the other two
$\eta$ problems.

Using~(\ref{V0con}), which follows from the COBE and spectral
index constraints (\ref{COBE}) and (\ref{nsconstr}), we find that
\begin{equation}
g x^2 e^{K/2} \lesssim 6 \times 10^{-5} \, .
\end{equation}
This is consistent with $g x^2 \ll 1$, as we have assumed
(and is required, for example, for $V_\mathrm{stab}$ to
retain its minimum during inflation).

Finally, for the maximal allowed value of $\tilde
V_0^{1/4}=2\times 10^{16}$ GeV, the relative contribution of
gravitational waves $r\approx 12.4 \epsilon$ is 13 percent.
However, it can be easily shown that the field variation
$\Delta\phi$ during inflation, related to the e-fold number by
\begin{equation}\label{Nfold}
N = \int \frac{V}{V'}d\phi \approx \frac{\tilde V_0}{C}\Delta\phi \, ,
\end{equation}
is larger than $M_{\rm Pl}$ (with $N=60$), as long as $\tilde V_0^{1/4}\gtrsim
7\times 10^{15}$ GeV. Since field variations below the Planck
scale are preferred theoretically, its very likely that the
predicted amount of gravitational waves is much lower.

\subsection{Moduli destabilisation and the gravitino mass}
\label{ssec:bar}

Here we study the stability of the modulus during inflation. In
particular, the previous approximation of assuming that the moduli
potential is quadratic when perturbed by the inflaton is only valid
if the inflaton does not remove the potential barrier between the
inflation minimum and infinity. We have to ensure that the term
linear in $\phi$ in the potential is not too big, otherwise this
term will completely destabilise the modulus $T$. The condition for
this to happen is that for at least 60 e-folds of inflation the term
linear in $\phi$ in~(\ref{VInf}) is smaller than the barrier height,
$V_\mathrm{max}$. Typically $V_\mathrm{max} \sim 3 W^2 e^K$ which
implies
\begin{equation}
2\sqrt{2}gx^2 e^K \tilde V_1\Delta \phi < V_\mathrm{max} \sim
3W^2 e^K \, .
\end{equation}
Using~(\ref{Nfold}) to eliminate $\Delta\phi$ from this
equation, we get
\begin{equation} \label{stabcond}
W \gtrsim \sqrt{N} W' T_0 \, ,
\end{equation}
where we have used ${\tilde V}_1 \approx -2 T_0W'$ (ignoring loop
corrections), and have dropped numbers of order one. Using the COBE
normalisation, we get
\begin{equation}
m_{3/2}  \gtrsim \sqrt{\frac{N}{\delta}} V_0
\end{equation}
where $m_{3/2}= W e^{K/2}$.  This is the condition guaranteeing
that the modulus is not destabilised during inflation. It states
that the gravitino mass is never small. In terms of the soft
breaking terms and the sparticle spectrum obtained with a large
gravitino mass, this implies that the scalar sparticles have very
large masses. This is the type of scalar spectrum advocated in
split SUSY models~\cite{splitsusy}. Of course if one is to observe
scalars at the LHC, the scalars cannot be very massive and the
gravitino mass must be reduced. 
This requires a superpotential which
gives a small gravitino mass, but which still produces a stabilisation
potential with a high barrier. The model in subsection~\ref{ssec:eg2} has
these properties. It is even possible to have $m_{3/2}=0$ and still
stabilise the moduli in a Minkowski vacuum, as described
in~\cite{kalloshlinde}. In this case there is no lifting term
$V_\mathrm{lift}$, and the vacuum is supersymmetric. We could then
obtain a gravitino mass from additional supersymmetry breaking effects. 
As we will show in section~\ref{sec:nogo}, if the vacuum is be
Minkowski, the SUSY-breaking must involve a lifting term. Otherwise the
vacuum will be unstable.

\section{KKLT scenario}
\label{sec:kklt}

As has already been noted, the parameters $\tilde V_0$, $C$ and
$M^2$ are not independent, but depend on details of the moduli
sector as well as on $g$ and $x$. As a simple example we will
discuss the KKLT scenario~\cite{KKLT}, which has just one modulus
field $T$ and uses a non-supersymmetric lifting term~(\ref{NSlift}).
During inflation the full potential is then given by~(\ref{kkltVS})
and (\ref{VInf})--(\ref{kkltV2}). The stabilisation potential's
minimum is at $T=T_0$, so $V_\mathrm{stab}'(T_0) =
V_\mathrm{stab}(T_0) =0$. This allows us to eliminate $E$, and one
other parameter.

Following~\cite{KKLT}, we will use a racetrack superpotential
\begin{equation}
W(T) = Ae^{-aT} + B e^{-bT} \, ,
\label{racetwo}
\end{equation}
where $B$, $A$, $a$ and $b$ are real constants. The above
superpotential comes from gaugino condensation. We take $T
\gg 1$ and $a T, b T \gg 1$ to guarantee the validity of the
supergravity approximation and the weak gauge coupling limit.

\subsection{Simple KKLT superpotential}
\label{ssec:eg1}

Let us first
focus on the KKLT model obtained with $b=0$ and $a T_0 \gg 1$.
To leading order this gives $W(T_0) \approx - (2/3)A a T_0 e^{-a T_0}$ and
\begin{eqnarray}
&&
\tilde V_1 \approx 2 A a T_0 e^{-aT_0}
\left[1+\frac{5\Lambda^2}{24\pi^2}\right]\, , \qquad
\tilde V_2 =-\frac{\Lambda^2}{4 \pi^2} \, ,
\nonumber \\ &&
(e^K\tilde V_1)'\approx -\frac{A a^2}{4T_0^2} e^{-aT_0}
\left[1+\frac{\Lambda^2}{8\pi^2}\right]  \, , \qquad
\lambda \approx \frac{A^2 a^4}{3T_0}e^{-2aT_0}
\left[1+\frac{\Lambda^2}{8\pi^2}\right] \, ,
\label{eg1}
\end{eqnarray}
where $\lambda = \tilde V''_\mathrm{stab}(T_0)$.

During inflation we also require the slow roll parameter $\epsilon
\approx [V_1/(g x^2)]^2 <1$. We see above that $V_1 \sim W$, and so
we need $W \lesssim g x^2$. We observe that $\tilde V_2$ is small,
suggesting $\eta$ is small too. However there are also corrections
to $\eta$ from the modulus field variation during inflation (see
section~\ref{sec:mod}), which cannot be ignored. These imply
\begin{equation}
\eta \approx
2\left(\tilde V_2 -2\frac{([e^K\tilde V_1]')^2}{e^K \lambda}\right) \, .
\label{neweta}
\end{equation}
We see that for the above choice of $F$-term model, the slow-roll parameter
is found to be negative, independent of the parameters, and
large: $\eta \approx -6.03$. This value is not compatible
with the slow-roll condition and, hence, not compatible with observational
constraints. In general there could also be corrections to $\eta$ from
variations of the imaginary part of $T$. However for the examples in
this section we find that the potential is always minimised at
$\Im (T)=0$, even during inflation.

Another constraint on the model comes from requiring that the
minimum of $\tilde V_\mathrm{stab}$ does not disappear during
inflation. This would be the case if $\tilde V_0 > V_\mathrm{max}$.
For the above example we find that the maximum of the
potential is at  $T_\mathrm{max} \approx T_0+ (1/a)\log(y a T_0)$, with
$y=2[1+\Lambda^2/(8\pi^2)]/[n+(5n-6)\Lambda^2/(24\pi^2)]$. The barrier
height is then
\begin{equation}
V_\mathrm{max} \approx \frac{1}{6T_0} A^2 a^2 e^{-2aT_0}
\left[1+\frac{\Lambda^2}{8\pi^2}\right] \, .
\end{equation}
We see that $V_\mathrm{max} \approx 3 W^2 e^K$, as we assumed in the previous
section.

We need $V_0 \sim g^2 x^4 e^K \lesssim V_\mathrm{max}$, implying $g
x^2 \lesssim W$. However this contradicts $g x^2 > W$, which was
required to have small $\epsilon$. Hence the moduli superpotential
in this subsection fails on both slow-roll parameters.

\subsection{Racetrack superpotential}
\label{ssec:eg2}

As a second case, consider the race-track models with $b \approx a$,
and take the limit $a T_0, b T_0 \gg 1$. In this case we find
$W(T_0) \approx - A(a-b) e^{-a T_0}/a$ and
\begin{eqnarray}
&&
\tilde V_1 \approx A(a-b) e^{-a T_0} \frac{3}{a}
\left[1+\frac{5\Lambda^2}{24\pi^2}\right] \, , \qquad
\tilde V_2 \approx -\frac{\Lambda^2}{4 \pi^2} \, ,
\nonumber \\ &&
(e^K \tilde V_1)' \approx -A(a-b) e^{-a T_0} \frac{a}{4T_0^2}
\left[1+\frac{\Lambda^2}{8\pi^2}\right]  \, ,
\nonumber \\ &&
\lambda \approx A^2 (a-b)^2 e^{-2aT_0} \frac{a^2}{3T_0}
\left[1+\frac{\Lambda^2}{8\pi^2}\right] \, .
\label{eg2}
\end{eqnarray}

Small $\epsilon$ requires $V_1 < g x^2$ which, as in the previous
subsection, implies $W < g x^2$. This suggests that it will again be
impossible to have small $\epsilon$ and a stable modulus $T$.

Using an asymptotic analysis of $\tilde V_\mathrm{stab}$, we find
that this time the maximum of the potential occurs at
$T_\mathrm{max} \approx T_0 +1/a$, and the height of the barrier
is
\begin{equation}
V_\mathrm{max} \approx \frac{1}{6T_0} A^2 (a-b)^2 e^{-2aT_0}
\left[1+\frac{\Lambda^2}{8\pi^2}\right] \, .
\end{equation}
In contrast to the previous subsection we find
$V_\mathrm{max}$ is not of order $W^2 e^K$. In this case it is
possible to have $\epsilon  \ll 1$ and $V_0 \lesssim V_\mathrm{max}$.

Unfortunately when we calculate $\eta$, we run into the same problem
as before. If the modulus field $T$ were fixed at $T_0$, $\eta$
would be small. However this does not occur, and~(\ref{neweta})
applies instead. We find again $\eta \approx -6.03$, ruling out this
case too.

\subsection{Analytic arguments}

We will now use approximate analytic arguments to determine if any
choices of $W$ could give viable slow-roll inflation. Using the
expressions for $\tilde V_\mathrm{stab}$~(\ref{kkltVS}), and
$\tilde V_1$~(\ref{kkltV1}), we obtain
\begin{eqnarray}
&& n \tilde V_\mathrm{stab} + T \tilde V'_\mathrm{stab}
=
\nonumber \\ && {}
-\frac{W'' W}{2T}
-\left[\frac{W''}{6T} + \frac{(2-n)}{4T^3}W\right] V_1
+\frac{(n-4)}{24 T^3} V_1^2
+ \mathcal{O}\left(\frac{\Lambda^2}{8 \pi^2}\right) = 0
\label{approx}
\end{eqnarray}
at $T=T_0$. Now the requirement that $\epsilon \ll 1$ implies that
$V_1 \ll g x^2$, and since $g x^2 \ll 1$ we can take $V_1 \ll 1$
to find approximate solutions to~(\ref{approx}). Noting that
$1/(8\pi^2) \approx 0.01$, the effect of the loop corrections
will also be small, allowing us to neglect them. For small
$V_1$~(\ref{approx}) is only satisfied if either $W$ or $W''$ is very
small. For the former we obtain
\begin{equation}
W = -\frac{V_1}{3}
+ \mathcal{O}\left(V_1^2, \frac{\Lambda^2}{8 \pi^2}\right) \, .
\end{equation}
Using this, the expression for $\lambda = \tilde
V''_\mathrm{stab}(T_0)$ simplifies to $\lambda = W''^2/(3T)$, to
leading order in $V_1$. From~(\ref{kkltV1}) we obtain
\begin{equation}
(e^K V_1)' =
- \frac{T W''-2W'}{4T^3} + \mathcal{O}\left(\frac{\Lambda^2}{8\pi^2}\right).
\label{dV1app}
\end{equation}
To leading order in $V_1$ this is $-W''/(4T^2)$.
Substituting all this into the expression for $\eta$~(\ref{neweta}),
which includes the effect of varying $T$, we find
\begin{equation}
\eta \approx -4 \frac{([e^K \tilde V_1]')^2}{e^K \lambda}
\approx -6 \, .
\end{equation}
So we retrieve an $\eta$ problem along this branch of solutions. The
models discussed in the previous two subsections fall into
this category.

The second approximate solution to~(\ref{approx}) has
\begin{equation}
W'' = \frac{(n-2)}{2 T_0^2} V_1
+ \mathcal{O}\left(V_1^2, \frac{\Lambda^2}{8 \pi^2}\right) \, .
\end{equation}
This again allows us to simplify $\lambda$, and we find
$\lambda \approx -W W'''/(2T^2)$. Expanding in $V_1$, (\ref{dV1app})
implies $(e^K V_1)' \approx -V_1/(8T^4)$. Combining all this,
(\ref{neweta}) implies
\begin{equation}
\eta \approx -\frac{V_1^2}{T_0^3 W'''W} \, .
\end{equation}
Noting that $\tilde V_1 \approx - 2 TW' \ll 1$ in order to get successful
inflation, we need both $T W'$ and $T^2 W'' \ll (T^3 W'''W)^{1/2}$
at the minimum $T=T_0$. If this fine-tuning is not satisfied then slow
roll inflation does not occur. This fine-tuning of the moduli
superpotential is in addition to the fine-tuning required for the
stabilisation potential $V_\mathrm{stab}$ to have a Minkowski
minimum. Furthermore, $W$ must also be such that the maximum of
$V_\mathrm{stab}$ is greater than the inflationary scale (so that
minimum of the moduli potential does not disappear). This all adds up
to a huge amount of fine-tuning, and the two-term racetrack potential
used above~(\ref{racetwo}) does not seem to have enough parameters to
satisfy all the conditions simultaneously. This suggests that a
heavily fine-tuned, three-term $W$ would be needed to give viable
inflation. It is hard to see how such a fine-tuned model could arise naturally.

\section{Supersymmetry Breaking Minkowski vacuum}
\label{sec:nogo}

In this section we consider Minkowski vacua with a non-vanishing
gravitino mass. This is the traditional framework used in particle
physics. At the end of inflation, the vacuum is a non-supersymmetric
configuration with a vanishing cosmological constant. One would also
like to impose $W\ll 1$ in order to have a hierarchy between the
Planck scale and the sparticle masses. Supersymmetry breaking could be
achieved with $F$-terms, although as we will show this leads to an
unstable vacuum. It is therefore necessary to include a lifting term
in the theory, such as those discussed at the end of
subsection~\ref{ssec:VV}.

Consider a theory with only one modulus field, no lifting term,
and the following no-scale Kahler potential
\begin{equation}
K=-3 \ln (T + \bar T) \, ,
\end{equation}
and $W=W(T)$. It is convenient to transform the field according to
\begin{equation}\label{kahlertrafo}
T=\frac{1-iz}{1+iz} \, .
\end{equation}
In terms of the new variable $z$, one obtains
\begin{equation}
K = -3 \ln (1- \vert z\vert^2)
\end{equation}
and
\begin{equation}
W(z)= \left(\frac{1+iz}{\sqrt 2}\right)^3 W\left(\frac{1-iz}{1+iz}\right) \, .
\end{equation}
The Kahler transformation~(\ref{kahlertrafo}) preserves
$G= K + \ln \vert W\vert^2$, which is the only relevant combination in
supergravity. Using the new variable $z$ one obtains
\begin{equation}
V(z)= \frac{1}{3(1-\vert z\vert ^2)^2}
(\vert W'\vert ^2 - \vert 3W - zW'\vert^2) \, ,
\end{equation}
where primes denote here $d/dz$. Imposing that the potential has an
extremum with a vanishing cosmological constant leads to the following
relations
\begin{equation}
\left| 3\frac{W}{W'} -z\right| =1
\end{equation}
 and
 \begin{equation}
\frac{W'^2}{WW''}= \frac{2}{3}
\end{equation}
at the extremum when $W\ne 0$ and $W'\ne 0$, i.e.\ when
supersymmetry is broken at the extremum. The extremum equations
are also satisfied for $W=W'=0$. The mass at the extremum involves
\begin{equation}
V_{z\bar z}= \frac{1}{3(1- \vert z \vert ^2)^2}(\vert W'' \vert^2
- \vert 2W'- zW''\vert^2)
\end{equation}
When $W\ne 0$ and $W'\ne 0$, the extremum equations give
\begin{equation}
 V_{z\bar z} =0
\end{equation}
at the extremum $z_0$. As $V_{zz}$ is generically non-zero, this
implies that the extremum is a saddle point where the direction
along $\Im(z-z_0)$ (for $V_{zz}$ chosen real positive) is
tachyonic. This is a generic property of no-scale models and
implies that the modulus cannot be stabilised with a vanishing
potential and broken supersymmetry.

On the contrary when supersymmetry is preserved, we find
\begin{equation}
V_{z\bar z}= \frac{ \vert W'' \vert^2}{3(1-\vert z \vert ^2)}>0 \
\end{equation}
as long as $W''$ does not vanish, guaranteeing the stability of
the model~\cite{stab}.

 This argument can be
generalised to an arbitrary number of fields in the no-scale case
\begin{equation}
K= -3\ln \left(T+ \bar T - \sum_i|\chi_i|^2 \right) \, .
\end{equation}
where the $\chi^i$'s are, for example,  matter fields.
 Indeed putting $z^i= (1+iz) \chi^i$ and
collectively $z^I=\{z,z^i\}$, the Kahler superpotential becomes
\begin{equation}
K=-3 \ln (1- z^I \delta_{IJ} \bar z^J)
\end{equation}
and the superpotential $W(z,\chi^i)$ is turned into
\begin{equation}
W(z^I)\equiv \left(\frac{1+iz}{\sqrt 2}\right)^3 W\left(\frac{1-iz}{1+iz},
\frac{z^i}{1+iz}\right).
\end{equation}
With these redefinitions, the scalar potential reads
\begin{equation}
V= \frac{1}{3(1-\vert z^I \vert^2)^2}  \left( \vert W_I\vert^2  -\vert 3W-
W_I z^I\vert^2 \right).
\end{equation}
The vanishing of the cosmological constant at the extremum implies
that
\begin{equation}
\vert W_I\vert^2  = \vert 3W- W_J z^J\vert^2 .
\end{equation}
 The minimum equation reads
\begin{equation}
(2W_I -z^JW_{JI})= \frac{W_{IJ}\delta^{J\bar K} \bar W_{\bar
K}}{3\bar W-\bar z^{\bar J}{\bar W_{\bar J}}}  \, .
\end{equation}
when supersymmetry is broken. In the supersymmetric case, the
minimum equations are $W=W_I=0$.

 The stability of the extremum
depends on the mass matrix which reads
\begin{eqnarray}
V_{I\bar J}&=&\frac{1}{3(1-\vert z^I \vert^2)^2}
\nonumber \\ && \hspace{0.3in} {} \times
\left[W_{KI} \delta^{K \bar K}\bar W_{\bar K\bar J}
- (2W_I - z^KW_{KI})(2\bar W_{\bar J}-\bar z^{\bar K}\bar
W_{\bar K \bar J})\right] \, .
\end{eqnarray}
Using the extremum equation we find that
\begin{eqnarray}
V_{I\bar J}&=&\frac{1}{3\vert W_I\vert ^2 (1-\vert z^I
\vert^2)^2}
\nonumber \\ && \hspace{0.3in} {} \times
\left(\vert W_I \vert^2W_{IK}\delta^{K\bar L}\bar W_{\bar L\bar J}-
W_{IL}\delta^{L\bar K}\bar W_{\bar K} \bar W_{\bar J\bar
M}\delta^{\bar M N}W_N\right) \, .
\end{eqnarray}
in the supersymmetry breaking case. This mass matrix has nice
algebraic properties. When $W_{IJ}$ has zero eigenvalues, the mass
matrix $V_{I\bar J}$ vanishes  in these directions. In the
directions orthogonal to the zero eigenstates of $W_{IJ}$,  all
the eigenvalues of $V_{I\bar J}$ are positive but one which
vanishes along the eigenvector $f^I=W^{IJ} W_J$ where
$W^{IJ}W_{JK}= \delta^I_K$.  Hence the mass matrix $V_{I\bar J}$
vanishes along $f^I$ and the zero eigenstates of $W_{IJ}$.

In the supersymmetric case, we find
\begin{equation}
V_{I\bar J}=\frac{1}{3(1-\vert z^I \vert^2)^2} {} \times
\left[W_{KI} \delta^{K \bar K}\bar W_{\bar K\bar J} -
z^KW_{KI}\bar z^{\bar K}\bar W_{\bar K \bar J})\right] \, .
\end{equation}
Defining a vector space basis with $e^I_0=z^I/\vert z^I\vert$
where $\vert z^I\vert ^2= z^I \delta_{I\bar I} \bar z^{\bar I}$,
and orthogonal vectors $e^I_i,\ i>0$, we can use the sum rule
$\delta^{K\bar K}= \sum _i e^K_i \bar e^{\bar K}_i$ to find that
\begin{eqnarray}
V_{I\bar J}&=&\frac{1}{3(1-\vert z^I \vert^2)^2} \nonumber \\ &&
\hspace{0.3in} {} \times \left[\left(\frac{1}{\vert z^I
\vert^2}-1\right)z^KW_{KI}\bar z^{\bar K}\bar W_{\bar K \bar J}-
\sum_{i>0} W_{IK}e^K_i \bar W_{\bar J \bar K} \bar e^{\bar
K}_i\right] \, .
\end{eqnarray}
which is a positive definite Hermitian matrix as $ V_{I\bar J}u^I
\bar u^{\bar J}>0$ for any $u^I$ provided the matrix $W_{IJ}$ does
not have zero eigenstates corroborating the results in~\cite{stab}.

Hence we have found that the mass matrix $V_{I\bar j}$ has zero
eigenstates when supersymmetry is broken. Now for a generic
supersymmetry breaking model, the holomorphic mass matrix is such
that $V_{IJ} f^I f^J \ne 0$. Expanding along the direction
$z^I=z^I_0 + \sigma f^I + \mathcal{O}(\sigma^2)$ where $z_0^I$ is
the minimum leads to a potential
\begin{equation}
V= \frac{1}{2} V_{IJ}f^If^J \sigma^2 + (\mathrm{c.c.})
+ \mathcal{O}(\sigma^3) \, .
\end{equation}
Notice that  this leads to a tachyonic direction as before,
implying that the supersymmetry breaking extremum with zero
cosmological constant is not stable. Hence we see that a
supersymmetry breaking Minkowski vacuum from a no-scale theory
which does not use a lifting term cannot be stable. Since such a
setup does not even stabilise the moduli fields, it is certainly
not suitable to use in a hybrid inflation model.

\section{Conclusions}\label{sec:conc}

In this paper we considered $F$-term hybrid inflation in supergravity, when
moduli fields are present. In the literature it is usually assumed
that moduli fields are stabilised before the period of slow-roll inflation
begins and that they therefore have no impact on the dynamics of the
inflaton field. We have shown that this is not necessarily the case. Firstly,
the tree-level coupling of moduli and inflaton fields in supergravity
induces a slope and, in general, a mass term in the potential for the
inflaton field. On top of this, loop corrections from both the
inflation and moduli sectors of theory contribute to the slope and curvature of
the potential. A further problem comes from the evolution of the moduli
fields during inflation. Since they are stabilised by a steep potential,
they are roughly constant during inflation. However, despite the small
size of their variation, we have shown that they still give a
significant contribution to the effective inflaton potential. This
generally leads to a large and tachyonic mass for the inflaton.

We find that this class of model has up to three different $\eta$
problems. Firstly there is the usual one coming from embedding hybrid
inflation in supergravity (which is avoided with the help of shift
symmetries). Secondly there are tree-level contributions to the
inflaton mass from the coupling of the inflaton to the moduli sector
(although these are absent for no-scale models). Finally there is the
negative contribution to the effective inflaton mass from the rolling
of the moduli fields during inflation. It is not obvious that
this third $\eta$ problem can be removed with natural additional
symmetries. The model also has an $\epsilon$ problem due to the
tree-level contributions to the inflaton slope from the moduli sector.

We found that in the popular model for moduli stabilisation, the KKLT
scenario, the induced mass term for the inflaton field is generally
too big to give a period of slow-roll inflation.
It seems that only with a heavily fine-tuned superpotential for the
moduli sector may viable slow-roll inflation take place.

To conclude, we have shown that the coupling of moduli fields to hybrid
inflation can have an important impact on the effective inflationary
potential. Neglecting this coupling is not consistent and, as we have
seen, generally leads to incorrect conclusions about the inflationary
dynamics. These considerations are also relevant to other inflationary
scenarios. In each case it should be checked that the moduli
fields, even if thought to be stabilised, do not spoil the flatness of the
inflationary potential, and therefore the dynamics of the inflaton field.

\ack
We would like to thank K. Dasgupta and R. Kallosh for comments on a
previous version of the manuscript and N. Chatillon and U. Ellwanger
for discussions. ACD thanks CEA Saclay for hospitality while some of
this work was in progress. ACD and CvdB were supported in part by
PPARC. SCD was supported  by the Swiss Science Foundation and the
Netherlands Organisation for Scientific Research (NWO). PhB
acknowledges support from RTN European programme MRN-CT-2004-503369.

\appendix
\section*{Appendix: Masses}
\setcounter{section}{1}

We have $M^2_{\rm (boson)} = 2 \mathcal{K}^{i\bar \jmath}
\partial_i \partial_{\bar \jmath} V + 6f_R^{-1} \sum_{i=\pm}
|\phi^i|^2$, where $f_R = \Re(f)$.
Note that we are assuming the only gauge group in the theory is
$U(1)_1$. If there were additional gauge groups, the additional
gauge sector would also give contributions to the mass matrices. For
the fermions
\begin{equation}
M^2_{\rm (fermion)} = 2 \mathcal{K}^{\bar lj} m_{ij}
\mathcal{K}^{\bar k i} \bar m_{\bar k \bar l}
+ 4 \mathcal{K}^{i\bar \jmath} m_{i\lambda}
\bar m_{\bar \jmath\lambda}
+ 2 |m_{\lambda\lambda}|^2 + 4 m^2_{3/2}
\end{equation}
where
\begin{eqnarray}
m_{ij} &=& e^{\mathcal{K}/2} D_i D_j \mathcal{W}
\nonumber \\
&=& e^{\mathcal{K}/2}\left( \mathcal{W}_{ij}+\mathcal{K}_{ij}\mathcal{W}
+ \mathcal{K}_i \partial_j \mathcal{W} + \mathcal{K}_j \partial_i \mathcal{W}
+ \mathcal{K}_i \mathcal{K}_j \mathcal{W}
- \Gamma_{ij}^k D_k \mathcal{W}\right)
\label{fermmod}
\end{eqnarray}
with $\Gamma_{ij}^k =\mathcal{K}^{k \bar l}\partial_i
\mathcal{K}_{j\bar l}$, and
\begin{equation}
m_{i \lambda}
= \frac{\sqrt{2} q_i \bar \phi^{\bar \imath}}{\sqrt{f_R}}
\, , \qquad
m_{\lambda\lambda} = \frac{e^{\mathcal{K}/2}}{2f_R}
 \mathcal{K}^{i \bar \jmath}D_{\bar \jmath} \bar \mathcal{W} \partial_i f
\, , \qquad
m_{3/2} = e^{\mathcal{K}/2}\mathcal{W} \, ,
\end{equation}
where $q_\pm =\pm 1$, and $q_i = 0$ otherwise.
Define
\begin{equation}
m^{(0)}_{IJ} = e^{K/2} D_I D_J W \, , \qquad
m^{(1)}_{IJ} = e^{K/2} [K_{IJ} +K_I K_J - \Gamma^K_{IJ} K_K] \, .
\end{equation}
The diagonal entries of the boson mass matrix for the Minkowski
background are
\begin{eqnarray}
M^2_{\pm \bar \pm} &=& (2g^2x^2 + 2|W|^2 + V_S) e^K
+ \frac{x^2}{f_R}\, ,
\nonumber  \\
M^2_{\phi\bar\phi} &=& (4g^2x^2 + 2|W|^2 + V_S) e^K  \, , \qquad
M^2_{I \bar J} = (e^K V_S + V_{\rm lift})_{,I\bar J} \, ,
\end{eqnarray}
and non-zero elements of fermion mass matrix are
\begin{eqnarray}
m_{\phi\phi} &=& m_{\pm \mp} = -m_{3/2} \, , \qquad
m_{\phi \pm} = e^{K/2}\sqrt{2} g x \, , \qquad
m_{\pm\lambda} = \pm \frac{\sqrt{2} x}{\sqrt{f_R}} \, , \qquad
\nonumber \\
m_{\lambda\lambda} &=&
e^{K/2} \frac{K^{I\bar J}D_{\bar J} \bar W \partial_I f}{2f_R} \, , \qquad
m_{IJ} =  m^{(0)}_{IJ} \, , \qquad
m_{3/2} = e^{K/2} W \, .
\end{eqnarray}
Hence we obtain
\begin{eqnarray}
\mathrm{Str} M^2_\mathrm{Mink} &=& 2 e^K |W|^2 + 6 e^K V_S + 2K^{I\bar J}
(e^K V_S +V_\mathrm{lift})_{,I\bar J}
\nonumber \\ && {}
- 2|m^{(0)}_{IK}K^{K\bar J}|^2
- \frac{e^K}{2f_R^2} |\partial_I f K^{I\bar J}D_{\bar J} \bar W|^2 \, .
\end{eqnarray}
For the inflationary background, the diagonal entries of the boson
mass matrix are
\begin{eqnarray}
M^2_{\phi\bar\phi} & = &
[2(3g^2x^4 +2g^2x^4\phi^2 -2\sqrt{2}gx^2\phi \Re(W) + |W|^2)
\nonumber \\ && \hspace{1in} {}
 + V_S -2\sqrt{2}gx^2\phi \Re (V_1) + 2g^2x^4(1+\phi^2)V_2]e^K \, , \
\nonumber \\
M^2_{\pm \bar \pm} & = & [2(g^2x^4 + g^2[2x^4+2x^2+1]\phi^2
-\sqrt{2}g \phi \Re(W)[1+2x^2] + |W|^2)
\nonumber \\ && \hspace{1in} {}
 + V_S - 2\sqrt{2}gx^2 \phi \Re (V_1)+2 g^2x^4\phi^2 V_2]e^K \, , \
\nonumber \\
M^2_{I \bar J} & = & 2g^2x^4 e^K (K_{I\bar J}+K_I K_{\bar J})
+ (e^K V_S + V_\mathrm{lift})_{,I\bar J}
\nonumber \\ && \hspace{1in} {}
- 2\sqrt{2}gx^2 \phi (e^K \Re V_1)_{,I\bar J}
+ 2 g^2 x^4 \phi^2 (e^K V_2)_{,I\bar J} \, ,
\end{eqnarray}
while fermion mass matrix has
\begin{eqnarray}
m_{\phi\phi} &=& -m_{3/2} \, , \qquad
m_{\pm \mp} = e^{K/2}(\sqrt{2} g \phi [1+ x^2] - W) \, ,
\nonumber \\
m_{\phi J} &=& -e^{K/2}\sqrt{2} g x^2 K_J \, , \qquad
m_{\lambda\lambda} =
e^{K/2} \frac{\partial_I f K^{I\bar J}}{2f_R}
(D_{\bar J} \bar W-\sqrt{2}g x^2 \phi K_{\bar J}) \, ,
\nonumber \\
m_{IJ} &=& m^{(0)}_{IJ} -\sqrt{2} g x^2\phi m^{(1)}_{IJ} \, , \qquad
m_{3/2} = e^{K/2} (W-\sqrt{2}g x^2 \phi) \, .
\end{eqnarray}
This implies
\begin{eqnarray}
\mathrm{Str} M^2_\mathrm{Inf} &=& \mathrm{Str} M^2_\mathrm{Mink} +
4\left[2+\delta^I_I\right]g^2x^4e^K
\nonumber \\ && {}
- 4\sqrt{2}g x^2\phi \Re \Biggl[(W+3 V_1) e^K
+ K^{I\bar J} (e^K V_1)_{,I\bar J}
\nonumber \\ && \hspace{0.3in} {}
- m^{(0)}_{IJ}K^{I\bar A} \bar m^{(1)}_{\bar A\bar B}K^{J\bar  B}
-\frac{e^K}{4 f_R^2}(K^{I\bar J} K_{\bar J} \partial_I f)
(K^{B\bar A}D_B W \partial_{\bar A} \bar f )
\Biggr]
\nonumber \\ &&  {}
+ 4g^2x^4\phi^2 \Biggl[(1+V_2) e^K + K^{I\bar J}(e^K V_2)_{,I\bar J}
\nonumber \\ && \hspace{1in} {}
- |m^{(1)}_{IK}K^{K\bar J}|^2
-\frac{e^K}{4f_R^2}|K^{I\bar J} K_{\bar J} \partial_I f|^2\Biggr] \, .
\end{eqnarray}

\end{document}